\documentclass[reprint,twocolumn,superscriptaddress,prl,showpacs,amsmath,amssymb,aps,showkeys]{revtex4-1}

\usepackage{amsfonts}
\usepackage{amsmath}
\usepackage{amssymb}
\usepackage[latin1]{inputenc}
\usepackage{graphicx}
\usepackage{dcolumn} 
\usepackage{bm}      
\usepackage{upgreek}
\usepackage{lmodern}
\usepackage{chemarrow}
\usepackage{mathtools}
\usepackage{upgreek}

\newcommand{\mr}[1]{\ensuremath{\mathrm{#1}}}

\newcommand{\ddx}[1][]{\ensuremath{\frac{\d #1}{\d x}}}
\newcommand{\ddxx}[1][]{\ensuremath{\frac{\d^2 #1}{\d x^2}}}

\newcommand{\B}{\mathrm{B}}

\renewcommand{\d}{\mathrm{d}}

\newcommand{\D}{\mathrm{D}}

\newcommand{\eo}{\mathrm{eo}}

\renewcommand{\H}{\mathrm{H}}

\newcommand{\ions}{\mathrm{ions}}

\newcommand{\mem}{\mathrm{mem}}

\newcommand{\p}{\mathrm{p}}
\renewcommand{\O}{\mathrm{O}}

\renewcommand{\r}{\mathrm{r}}
\newcommand{\res}{\mathrm{res}}
\newcommand{\R}{\mathrm{R}}

\newcommand{\RH}{\mathrm{RH}}

\newcommand{\sdl}{\mathrm{sdl}}

\newcommand{\T}{\mathrm{T}}
\newcommand{\tot}{\mathrm{tot}}
\newcommand{\w}{\mathrm{w}}

\newcommand{\HHO}{\ensuremath{{\H_2\O}}}

\newcommand{\Hp}{\ensuremath{{\H^+}}}
\newcommand{\Kw}{\ensuremath{{K_\w}}}
\newcommand{\pH}{\ensuremath{{\p\H}}}
\newcommand{\pK}{\ensuremath{{\p K}}}
\newcommand{\OH}{\ensuremath{{\O\H}}}
\newcommand{\OHm}{\ensuremath{{\OH^{-}}}}

\newcommand{\eps}{\ensuremath{\epsilon}}

\newcommand{\kBT}{\ensuremath{k_\mathrm{B}T}}

\newcommand{\veps}{\ensuremath{\varepsilon}}

\newcommand{\beq}[1]{\begin{equation} \eqlab{#1}} 
\newcommand{\eeq}{\end{equation}}
\newcommand{\bsub}{\begin{subequations}} 
\newcommand{\esub}{\end{subequations}}   
\def\bal#1\eal{\begin{align}#1\end{align}}
\def\bsubal#1\esubal{\bsub \begin{align}#1\end{align} \esub}

\newcommand{\eqlab}[1]{\label{eq:#1}}
\renewcommand{\eqref}[1]{Eq.~(\ref{eq:#1})}

\newcommand{\figref}[1]{Fig.~\ref{fig:#1}}

\newcommand{\figlab}[1]{\label{fig:#1}}

\begin{document}

\title{Current-Induced Membrane Discharge}


\author{M. B. Andersen}
\affiliation{Department of Micro- and Nanotechnology, Technical University of Denmark, DTU Nanotech Building 345 East, DK-2800 Kongens Lyngby, Denmark}
\affiliation{Department of Mechanical Engineering, Stanford University, Stanford, CA, 94305, USA}

\author{M. van Soestbergen}
\affiliation{Wetsus, Centre of Excellence for Sustainable Water Technology, Agora 1, 8934 CJ Leeuwarden, The Netherlands}
\affiliation{Department of Applied Physics, Eindhoven University of Technology, Den Dolech 2, 5612 AZ Eindhoven, The Netherlands}

\author{A. Mani}
\affiliation{Department of Mechanical Engineering, Stanford University, Stanford, CA, 94305, USA}

\author{H. Bruus}
\affiliation{Department of Micro- and Nanotechnology, Technical University of Denmark, DTU Nanotech Building 345 East, DK-2800 Kongens Lyngby, Denmark}
\author{P. M. Biesheuvel}

\affiliation{Wetsus, Centre of Excellence for Sustainable Water Technology, Agora 1, 8934 CJ Leeuwarden, The Netherlands}
\affiliation{Department of Environmental Technology, Wageningen University, Bornse Weilanden 9, 6708 WG Wageningen, The Netherlands}
\author{M. Z. Bazant}

\affiliation{Departments of Chemical Engineering and Mathematics, Massachusetts Institute of Technology, Cambridge, MA, 02139, USA.}

\date{\today}

\pacs{47.57.jd, 87.16.dp, 82.45.Mp, 82.33.Ln}


\begin{abstract}
Possible mechanisms for over-limiting current (OLC) through aqueous ion-exchange membranes (exceeding diffusion limitation) have been debated for half a century. Flows consistent with electro-osmotic instability (EOI) have recently been observed in microfluidic experiments, but the existing theory neglects chemical effects and remains to be quantitatively tested. Here, we show that charge regulation and water self-ionization can lead to OLC by ``current-induced membrane discharge" (CIMD), even in the absence of fluid flow. Salt depletion leads to a large electric field which expels water co-ions, causing the membrane to discharge and lose its selectivity. Since salt co-ions and water ions contribute to OLC, CIMD interferes with electrodialysis (salt counter-ion removal) but could be exploited for current-assisted ion exchange and pH control.  CIMD also suppresses the extended space charge that leads to EOI, so it should be reconsidered in both models and experiments on OLC. 
\end{abstract}

\maketitle


Selective ion transport across charged, water-filled membranes plays a major role in ion exchange and desalination~\cite{helfferich_book,Tongwen2005}, electrophysiology~\cite{weiss_book}, fuel cells~\cite{ohare_book,berg2004},  and lab-on-a-chip devices~\cite{Wang2005, Kim2007, Schoch2008, Sparreboom2009, Kim2010a}, but is not yet fully understood.  A long-standing open question has been to explain  experimentally observed overlimiting current (OLC), exceeding classical diffusion limitation~\cite{Levich1962}. Possible mechanisms include electroosmotic instability (EOI) and water splitting in the bulk solution~\cite{Nikonenko2010,Cheng2011}, as well as surface conduction and electro-osmotic flow in microchannels~\cite{Dydek2011}. Vortices consistent with EOI have recently been observed under OLC conditions ~\cite{Rubinstein2008, Yossifon2008, Kim2007}, although the theory of Rubinstein and Zaltzman~\cite{Rubinstein2000, Rubinstein2002, Zaltzman2007} remains to be tested quantitatively. The water splitting mechanism, either catalyzed by membrane surface groups or through the second Wien effect, has not yet been conclusively tied to OLC~\cite{Simons1979a, Simons1984, Danielsson2009, Tanaka2010,Cheng2011}.

In this Letter, we propose a chemical mechanism for OLC, ``current-induced membrane discharge" (CIMD), resulting from membrane (de)protonation and water self-ionization, even in the absence of fluid flow. The amphoteric nature of the charge of ion-exchange membranes (i.e. sensitivity to pH and other stimuli) is well known \cite{Ramirez1997, Ramirez1997a, Lint2002, Takagi2003, Bandini2005, Jensen2011,berg2004}, but not the response to a large applied current. The basic physics of CIMD is illustrated in Fig.~\ref{fig:domain} for an anion-exchange membrane. During OLC, a large electric field develops on the upstream, salt-depleted side of the membrane, which expels H$^+$ and attracts OH$^-$, causing the membrane  to deprotonate and lose selectivity, thereby allowing salt co-ions to pass and producing large pH gradients. The upstream solution becomes more acidic (low pH), while the downstream, salt-enriched solution and the membrane become more basic (high pH).

\begin{figure}[!t]
    \includegraphics[]{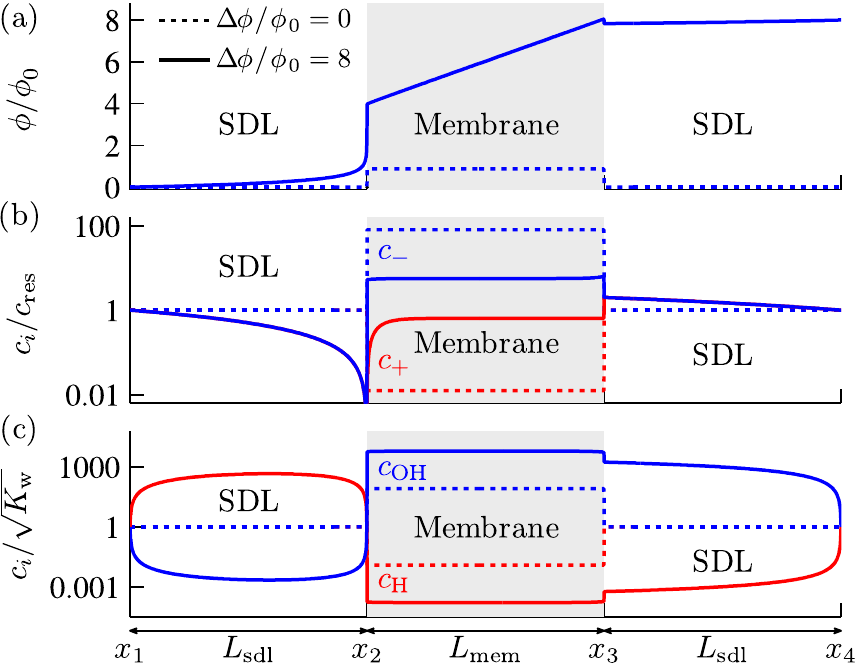}
    \caption{\figlab{domain} [Color online] Basic physics of CIMD, illustrated by numerical solutions of Eqs. (\ref{eq:alpha}), (\ref{eq:NP}), and (\ref{eq:Poisson}) for an anion exchange membrane between two stagnant diffusion layers (SDL) for (a) electrostatic potential and concentrations of (b) cations $c_+$ and anions $c_-$ and (c) protons $c_\H$ and hydroxyl ions $c_\OH$. }
\end{figure}

The local charge of an aqueous membrane strongly depends on the local pH. In our examples below, we consider an anion-exchange membrane with fixed surface groups of volumetric concentration $c_\mem$, which selectively allows negatively charged anions (counter-ions) to pass, while blocking cations (co-ions)~\cite{Yaroshchuk2012}. Depending on $\pH \approx \mbox{p[H]} = -\log_{10}(c_\H)$, where $c_\H$ is the proton concentration (H$^+$ or H$_3$O$^+$) in M, the membrane can ``discharge" (deprotonate): 
\begin{equation}
\RH^+ \xleftrightharpoons{K} \R + \Hp.   \label{eq:RH}
\end{equation}
The ratio of product to reactant concentrations in equilibrium is the dissociation constant $K$ in M ($\pK=-\log_{10}K$). Assuming a classical Langmuir adsorption isotherm \cite{Ramirez1997, Ramirez1997a, Lint2002, Takagi2003, Bandini2005, Koehler2006, Biesheuvel2006}, the ionization degree of the membrane, 
\bal\eqlab{alpha}
    \alpha &= \left(1+ \frac{K}{c_\H}\right)^{-1} = \left(1 + 10^{\pH - \pK}\right)^{-1},
\eal
relates its charge concentration $\alpha\, c_\mem$ to $\pH$ and $\pK$. 
(For a cation-exchange membrane, the power is $\pK - \pH$.) To describe the local pH, we cannot assume Boltzmann equilibrium with an external reservoir. Instead, we consider ion transport coupled to membrane discharge Eq.~(\ref{eq:RH}) and water self-ionization,
\begin{equation}
\HHO \xleftrightharpoons{\Kw} \OHm + \Hp,  \label{eq:HHO}
\end{equation}
with dissociation constant
\begin{equation}
K_\w = c_\H c_\OH    \label{eq:Kw}
\end{equation}
where $K_\w=10^{-14}$~M$^2$ at $T = 25~^\circ$C.  Although kinetics can be included ~\cite{Simons1979a,Simons1984,Biesheuvel2002,Danielsson2009,berg2004}, the reactions (\ref{eq:RH}) and (\ref{eq:HHO}) are typically fast, so we assume local quasi-equilibrium. 

We now develop a membrane model (seemingly the first) including all of these effects: (i) transport of four ionic species, including co-ions and water ions (\Hp{} and \OHm{}) along with majority anions, (ii) water self-ionization, and (iii) pH-dependent membrane charge. We consider the prototypical 1D electrodialysis geometry in Fig.~\ref{fig:domain}, consisting of a planar ion-selective membrane of thickness $L_\mem$ between two well-stirred reservoir compartments of salt ion concentration $c_\res$ and pH of $\pH_\res$. We adopt the simplest and most commonly used model of diffusion limitation~\cite{Levich1962}, in which ion concentrations vary across ``stagnant diffusion layers" (SDL) of thickness $L_\sdl$ (of the order 10--100~$\upmu$m) between the reservoirs and the membrane, e.g. representing convection-diffusion boundary layers or stagnant gel films. 
 
 Ionic diffusion, electromigration and reactions are described by four Nernst-Planck equations.   Following Refs.~\cite{Ramirez1997, Ramirez1997a, VanSoestbergen2010}, we  combine the Nernst--Planck equations for \Hp{} and \OHm{} using Eq.~(\ref{eq:Kw}) to eliminate the reaction terms and relate the water-ion current density $J_\w$ to the water-ion variable $c_\w = (D_\H c_\H - D_\OH c_\OH)/D_\w$, in which $D_\w = \sqrt{D_\H D_\OH}$ is the geometric mean of the free  \Hp{} and \OHm{} diffusivities. We thus arrive at the following set of coupled, nonlinear, differential equations to be solved in both SDLs and the membrane~\cite{VanSoestbergen2010}:
\bsub\eqlab{NP}
\begin{alignat}{2}
    \ddx[J_i] &=0,& i&=+,-,\w,\eqlab{diff_eqs1} 
    \\
    J_\pm &= \mp f_\r D_\pm &\Big(&\ddx[c_\pm] \pm c_\pm \ddx[\phi]\Big),
    \\
    J_\w &= - f_\r D_\w &\Big(&\ddx[c_\w] + \left[4\Kw + c_\w^2\right]^{\frac{1}{2}}\ddx[\phi]\Big),\eqlab{j_rho}
\end{alignat}
\esub
where $J_i$ is the ionic current density of species $i$ and $f_\r$ is a hindrance factor accounting for porosity, tortuosity and constriction ($f_\r=1$ in the SDLs). Here, $\phi$ is the dimensionless mean electrostatic potential scaled to the thermal voltage $V_\T = \kBT/e = 25.7$~mV and satisfying Poisson's equation
\bal\eqlab{Poisson}
    \ddxx[\phi] &= -4\,\pi\lambda_\B\left(\rho_\ions + \rho_\mem\right),
\eal
where $\lambda_\B = e^2/(4\,\pi\veps_{\r,j}\veps_0\kBT)$ is the Bjerrum length, and $\rho_\ions = \eps\,\left(c_+ - c_- + c_\H - c_\OH\right)$ and $\rho_\mem = \alpha\,\eps\,c_\mem$ are charge densities due to the ions and the immobilized charges in the membrane, respectively. The porosity $\eps$ of the membrane appears because concentrations $c_i$ are defined with respect to the interstitial, not total, volume ($\eps=1$ in the SDLs). In our simulations below, we choose the following typical parameters: $c_\mem = 5$~M, $\pK = 9.5$, $L_\mem = L_\sdl = 100~\upmu$m, $\veps_{\r,\sdl} = 78$, $\veps_{\r,\mem} = 29$, $\eps = 0.4$, $f_\r = 0.02$~\cite{Elattar1998}, $D_+ = 1.3\times 10^{-9}$~m$^2$\,s$^{-1}$ and $D_- = 2.0\times 10^{-9}$~m$^2$\,s$^{-1}$ (corresponding to NaCl), $D_\H = 9.3\times 10^{-9}$~m$^2$\,s$^{-1}$, and $D_\OH = 5.3\times 10^{-9}$~m$^2$\,s$^{-1}$. 
We also use $\pH_\res = 7$ and $\beta = 2\,c_\res/c_\mem = 0.02$, unless otherwise noted. The voltage difference across the system is $\Delta \phi$. At the reservoir/SDL boundaries we set $c_\pm = c_\res$ and relate $c_\w$ to $\pH_\res$.

In spite of neglecting fluid flow, the model still predicts OLC, as shown in \figref{domain}. The classical ion concentration polarization phenomenon is apparent in panel (b) with salt depletion where counter-ions (anions) enter ($x=x_2$) and enrichment where they leave ($x=x_3$). Within the membrane, however, anion depletion and cation (co-ion) enrichment reveal a significant loss of selectivity due to CIMD. At the same time, panel (c) shows large, order-of-magnitude variations in $c_\H$, ``mirrored" by $c_\OH$ in equilibrium Eq.~(\ref{eq:Kw}), with proton enrichment (acidity) in the left SDL and proton depletion (basicity) in both the membrane and the right SDL. The existence of such pH variations has been confirmed experimentally in similar systems \cite{Thoma1977, Jialin1998, Krol1999, Choi2001}.

Motivated by this observation, we analyze the pH gradients perturbatively in the full CIMD model. We consider under-limiting currents, assume thin, quasi-equilibrium double layers (Donnan approximation) at the SDL/membrane interfaces, and solve the leading-order problem for $c_+$, $c_-$ and $\phi$  with small perturbations in $c_\H$ and $c_\OH$, valid when $(c_\H-c_\OH)/(c_+-c_-)\ll1$. The resulting semi-analytical model (to be described in detail
elsewhere) suffices to predict CIMD (variations of membrane charge with local pH) via~\eqref{alpha}. Numerical calculations show that pH and $\alpha$ are nearly constant across the membrane,  so the water charge density is averaged between positions $x_2$ and $x_3$  (see below) to calculate the membrane charge and midplane pH [\figref{c_H}(b)] to be used in ~\eqref{alpha} to calculate $\alpha$.  

The final result for the most general model including membrane discharge, arbitrary values for $\pH_\res$ and $c_\res$, and the possibility that all diffusion coefficients are different, consists of~\eqref{alpha} together with the set of algebraic equations below (see the Supplemental Material for details). First, we introduce the dimensionless salt flux variable $j_\mr{salt} = (J_- - J_+\,D_-/D_+ )/J_{\lim}$, in which $J_{\lim} = -2\, D_- c_\res/L_\sdl$ is the ``classical'' limiting current density~\cite{Levich1962}, and obtain the salt current-voltage relation,
\bal\eqlab{jion}
	\Delta \phi = 4\,\tanh^{-1}\left(j_\mr{salt}\right) + \frac{j_\mr{salt}}{\gamma}\frac{\beta}{\alpha},
\eal
in which $\gamma = f_\r/l_\mem$, where $l_\mem = L_\mem/L_\sdl$ is the membrane-to-SDL width ratio.  The first term describes concentration polarization in the SDLs, while the second is the Ohmic response of the membrane. Next, we introduce the dimensionless water ion flux $j_\w = J_\w\,L_\sdl/(D_\w \sqrt{K_\w})$ and water ion variable $\rho_\w = c_\w/\sqrt{K_\w}$ and obtain the following equations, $\rho_\w(x_3^\mem) - \rho_\w(x_2^\mem)\exp[j_\mr{salt}\,\beta/(\gamma\,\alpha)] + j_\w/\gamma = 0$, $\sinh^{-1}[\rho_\w(x_i^\mem)/2] = \sinh^{-1}\left[\rho_\w(x_i^\sdl)/2\right] - \sinh^{-1}(\alpha/[\beta\,(1 \mp j_\mr{salt})])$, and $\rho_\w(x_i^\sdl) = \rho_\w^\res \mp \, j_\w + \rho_0[1 + 2\,\gamma\,\beta/\alpha]\ln(1 \mp\,j_\mr{salt})$, (where in these expressions $i = 2$ and 3 corresponds to $-$ and $+$, respectively). Here, $\rho_\w^\res$ is related to $\pH_\res$ and $\rho_0 = [4 + (\rho_\w^\res)^2]^\frac{1}{2}$. Note that $x_i^\mem$ and $x_i^\sdl$ refer to positions on either side of the equilibrium electric double layer at the membrane-SDL interfaces. In the limit of an infinite membrane charge $\beta/\alpha \to 0$ the solution to the leading order problem [\eqref{jion}] is simply the ``classical'' result~\cite{Bazant2005}, $j_\mr{salt} = \tanh\left(\Delta \phi/4\right)$. We find the characteristic voltage factor $\phi_0$ by expanding~\eqref{jion} for small $j_\mr{salt}\ll1$ and obtain $j_\mr{salt} = \Delta \phi/\phi_0$ in which $\phi_0 = 4 + \beta/\gamma$ assuming constant $\alpha = 1$. 

\begin{figure}[!tb]
    \includegraphics[]{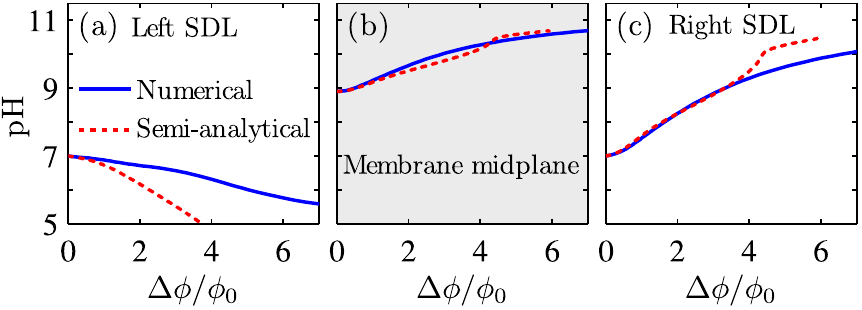}
    \caption{\figlab{c_H}[Color online] Predicted pH variations from the full numerical model, compared to the semi-analytical approximation, as a function of the applied voltage (a) in the left SDL, just next to the membrane, (b) at the membrane midplane, and (c) in the right SDL, next to the membrane. }
\end{figure}

Results of the semi-analytical model are compared with full numerical calculations in \figref{c_H}, which shows good agreement in the expected range of validity $\Delta \phi/\phi_0 \lesssim 1$. The pH appears to converge towards a limiting value for $\Delta \phi \to\infty$, and the jump in this limiting pH-value between the left SDL and the membrane is huge, here about 5 pH units at the highest values of $\Delta \phi$ considered. We note that the deviation between the analytical and numerical solution is largest in the left SDL where electroneutrality is most strongly violated. This comparative analysis constitutes a validation of our numerics and provides further support for our conclusions regarding the role of pH as controlling the ionic transport properties of ion-selective membranes.

\begin{figure}[!tb]
    \includegraphics[]{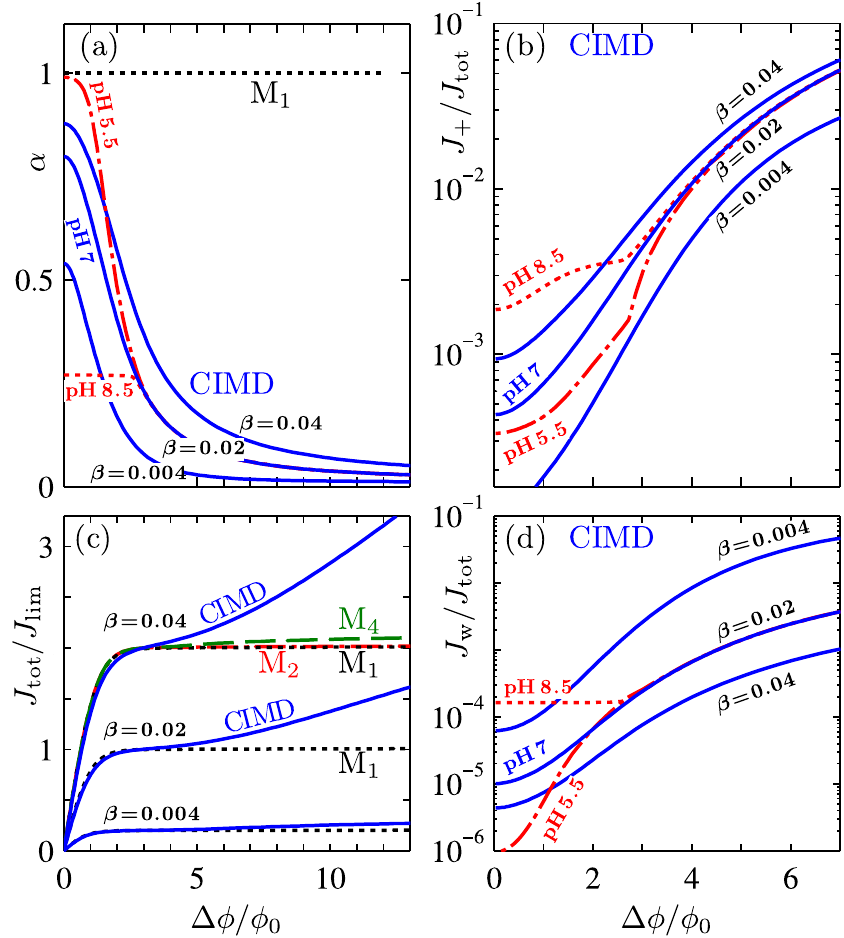}
    \caption{\figlab{IV} [Color online] Comparison of the classical M$_1$ model (only counterions in the membrane) with the full CMID model (label ``pH'' refers to $\pH_\res$). (a) Membrane ionization degree $\alpha$. (b) co-ion current $J_+$. (c) Total current $J_\tot$. (d) Water ion current $J_\w$.}
\end{figure}

We now turn to a numerical analysis of the CIMD model Eqs.~(\ref{eq:alpha})--(\ref{eq:Poisson}). For comparison, we also solve the classical model M$_1$ used in all prior work on EOI~\cite{Rubinstein2000, Rubinstein2002,  Zaltzman2007, Rubinstein2008, Yossifon2008} in which (i) $c_-=0$ in the membrane, (ii) $c_\H=c_\OH=0$ everywhere, and (iii) $\alpha=1$ for all conditions. We also solve two intermediate models which include co-ions in the membrane with $\alpha = 1$ and either \textit{exclude} (M$_2$) or \textit{include} (M$_4$) water ions, i.e.\ taking 2 or 4 ions into account in the membrane, respectively. The total current density is $J_\tot = J_+ + J_- + J_\w$.

Figure~\ref{fig:IV}(a) shows the significant decrease in the ionization degree $\alpha$ predicted by the CIMD model, in contrast to the constant $\alpha = 1$ in the M$_1$ model. Moreover, $\alpha$ decreases with decreasing $\beta$ (due to increasing Donnan potential) and decreases with $\pH_\res$ (due to decreasing $c_\H$ in the membrane). A striking, and yet unexplained prediction is that for $\pH_\res$ larger than 7 the ionization degree is almost constant until the curve hits that for $\pH_\res = 7$ after which the curves follow each other. In general we find beyond a few times $\phi_0$ that reservoir pH has a very small influence on membrane charge, fluxes and currents (see also~\figref{IV}(b)-(d)). Figure~\ref{fig:IV}(b) shows the significant increase of co-ion flux $J_+$, thus loss of membrane selectivity, with increasing voltage, as predicted by the CIMD model, for all values of pH and $\beta$, while \figref{IV}(d) shows likewise the increase in current density $J_\w$ due to water ions. Still, these contributions do not sum to the increased current during OLC, as shown in \figref{IV}(c), the difference being due to increased counter-ion flux $J_-$.

Although the current-voltage relation in CIMD is quite complicated, our simulations and analysis suggest two general trends: (i)  OLC increases with reservoir salt concentration,  roughly as $\beta^{0.65}$ for the parameters of Fig. 3; (ii) OLC is nearly independent of reservoir pH, in spite of the large pH gradients produced across the membrane. 

Finally, we analyze the possible effect of CIMD on EOI. In the classical M$_1$ model, non-equilibrium space charge forms at the limiting current ~\cite{Rubinstein1979,Bazant2005,Chu2005,Biesheuvel2009}, and its growing separation from the membrane reduces viscous resistance to electro-osmotic flow and destabilizes the fluid~\cite{Rubinstein2000,Rubinstein2002,Zaltzman2007}.  As a measure of the propensity to develop EOI we use the transverse (Helmholtz--Smoluchowski) electroosmotic mobility $\mu_{\eo}/\mu_{\eo,0}$ at the left SDL-reservoir edge, which is equal to the first moment of the charge density, $-4\,\pi\lambda_\B\,\int_{x_1}^{x_2}x\,\rho_\ions\,\d x$, or the dimensionless potential difference across the left SDL, $\phi(x_1)-\phi(x_2)$.

\begin{figure}[!tb]
    \includegraphics{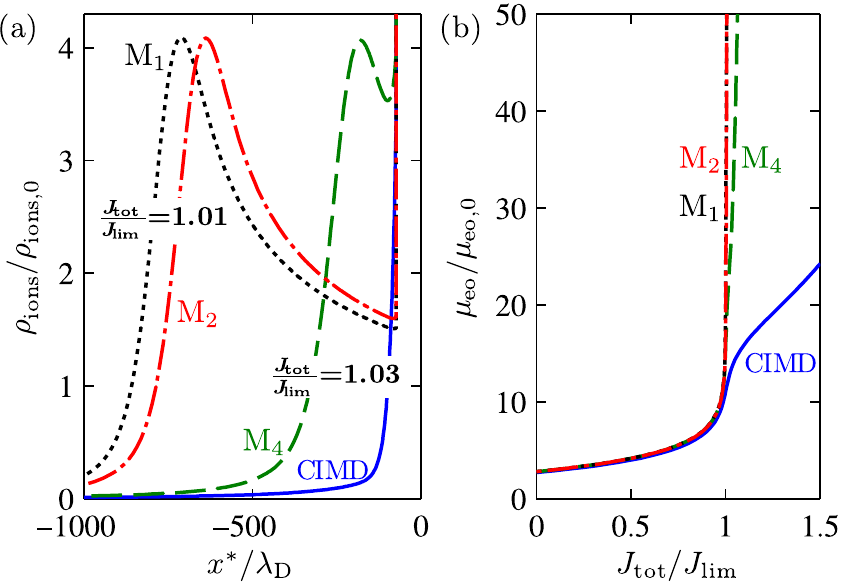}
    \caption{\figlab{fields}[Color online] Comparison of three fixed-charge membrane models $M_n$ having $n=$ 1, 2, or 4 mobile ionic species and the CIMD model for (a) charge density $\rho_\ions$ versus distance $x^*$ from the membrane scaled to the reservoir Debye length $\lambda_\D$, and (b) electroosmotic mobility $\mu_\eo$ as function of total current $J_\tot$.}
\end{figure}

Figure~\ref{fig:fields}(a) shows that slightly above the limiting current ($J_\tot/J_{\lim} = 1.01$) the  M$_1$ model already predicts a very significant extended space charge layer (the ``shoulder'' maximum in $\rho_\ions/\rho_{\ions,0} = (\lambda_\D\,L_\sdl/\phi_0)\,\d^2\phi/\d x^2$ several hundred Debye lengths from the membrane), whereas for an even higher current ($J_\tot/J_{\lim} = 1.03$), using the more realistic CIMD model, the extension of this layer is still very minor. The two intermediate models lie in between. Figure~\ref{fig:fields}(b) shows how the transverse electroosmotic mobility is predicted by the M$_1$ model to diverge at the limiting current. This divergence is significantly reduced only by the full CIMD model including simultaneously co-ion access, water ion transport, water splitting, and membrane discharge. We note that a proper analysis of EOI would be more involved, since here we have simply focused on the transverse electroosmotic mobility as a way of illustrating the suppression of EOI due to CIMD.

In conclusion, we have theoretically demonstrated that OLC through aqueous ion-exchange membranes can result from CIMD, or loss ion selectivity due to \mbox{(de-)protonation} coupled to ion transport and water self-ionization. The appearance of OLC carried partially by salt co-ions and water ions reduces separation efficiency in electrodialysis, but the associated large pH gradients and membrane discharge could be exploited for current-assisted ion exchange or pH control.  CIMD also suppresses the non-equilibrium space charge responsible for EOI and thus should be considered in both models and experiments on OLC with fluid flow. Although we have developed the theory for ion-exchange membranes in aqueous solutions, CIMD could occur in any nanofluidic system with an electrolyte whose ions regulate the surface charge.



%

\end{document}